\documentclass[12pt]{article}
\usepackage[dvips]{graphicx}
\usepackage{epsfig}

\def\fund{  \> {\vcenter  {\vbox
               {\hrule height.6pt
                \hbox {\vrule width.6pt  height5pt
                      \kern5pt
                      \vrule width.6pt  height5pt}
                \hrule height.6pt}
                         }
               }
            \>\>  }

\def\afund{  \> \overline{ {\vcenter  {\vbox
               {\hrule height.6pt
                \hbox {\vrule width.6pt  height5pt
                      \kern5pt
                      \vrule width.6pt  height5pt}
                \hrule height.6pt}
                         }
               } }
            \>\>  }

\def\sym{  \> {\vcenter  {\vbox
              {\hrule height.6pt
               \hbox {\vrule width.6pt  height5pt
                      \kern5pt
                      \vrule width.6pt  height5pt
                      \kern5pt
                      \vrule width.6pt  height5pt}
               \hrule height.6pt}
                         }
               }
            \>\>  }

\def\symbar{  \> \overline{ {\vcenter  {\vbox
              {\hrule height.6pt
               \hbox {\vrule width.6pt  height5pt
                      \kern5pt
                      \vrule width.6pt  height5pt
                      \kern5pt
                      \vrule width.6pt  height5pt}
               \hrule height.6pt}
                         }
               } }
            \>\>  }

\def\asym{ \> {\vcenter  {\vbox
                 {\hrule height.6pt
                  \hbox {\vrule width.6pt  height5pt
                         \kern5pt
                         \vrule width.6pt  height5pt}
                  \hrule height.6pt
                  \hbox {\vrule width.6pt  height5pt
                         \kern5pt
                         \vrule width.6pt  height5pt}
               \hrule height.6pt}
                         }
               }
            \>\>  }

\def\asymbar{ \> \overline{ {\vcenter  {\vbox
                 {\hrule height.6pt
                  \hbox {\vrule width.6pt  height5pt
                         \kern5pt
                         \vrule width.6pt  height5pt}
                  \hrule height.6pt
                  \hbox {\vrule width.6pt  height5pt
                         \kern5pt
                         \vrule width.6pt  height5pt}
               \hrule height.6pt}
                         }
               } }
            \>\>  }

\begin{document} 

\baselineskip 6.2mm 
\begin{center}
{\Large \bf
Color Superconductivity via Supersymmetry\footnote{Talk 
 presented by N.M. at YITP workshop on Fundamental Problems and Applications 
 of Quantum Field Theory, December 16-18, 2004, 
 Yukawa Institute for Theoretical Physics, Kyoto, Japan 
 and at 2004 International Workshop on Dynamical Symmetry Breaking, 
 December 21-22, 2004, Nagoya University, Nagoya, Japan.}
}
\end{center}
\begin{center}
{\large Nobuhito Maru\footnote{e-mail: 
maru@riken.jp,} and Motoi Tachibana\footnote{e-mail: motoi@riken.jp}
}
\end{center}
\begin{center} 
Theoretical Physics Laboratory, RIKEN \\
2-1 Hirosawa, Wako, Saitama, 351-0198, JAPAN \\
\end{center}
\abstract{
In this talk, 
a supersymmetric (SUSY) composite model of color superconductivity is discussed. 
In this model, quark and diquark supermultiplets are dynamically generated 
as massless composites by a newly introduced confining gauge dynamics. 
It is analytically shown that the scalar component of diquark supermultiplets 
develops vacuum expectation value (VEV) at a certain critical chemical potential. 
We believe that our model well captures 
aspects of the diquark condensate behavior 
and helps our understanding of its dynamics in real QCD. 
The results obtained here might be useful when we consider a theory 
composed of quarks and diquarks. 
}

\normalsize\baselineskip=15pt

\section{Introduction}
\begin{figure}[b]
\begin{center}
\includegraphics[width=7cm,height=4cm]{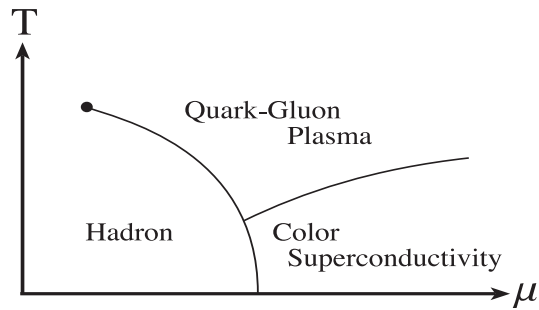}
\end{center}
\caption{\label{fig:phase} 
(Conjectured) phase diagram of hot and dense quark matter.}
\end{figure}

There has been much attention to color superconductivity conjectured 
to appear in QCD at high baryon density \cite{CSC}. 
Conjectured phase diagram is shown in Fig \ref{fig:phase}. 
Previous studies of QCD with this direction 
were limited to very high density 
where the theory is perturbative because of the asymptotic freedom \cite{son}. 
But our interest is the low energy physics 
where the theory is strongly coupled. 
Therefore, it would be interesting to find QCD-like theories 
calculable even at lower density, 
namely in the strong coupling regime. 
In this viewpoint, Nambu-Jona-Lasino model has been well studied 
by many people \cite{NJL}.

Recently, another approach using a SUSY gauge theory was proposed \cite{HLM}, 
where the breaking pattern of global symmetries in a softly broken SUSY QCD 
at finite chemical potential was investigated 
and compared to that obtained from the analysis via
nonsupersymmetric QCD \cite{schafer}. 
In particular, the phase structure associated with
baryon number symmetry $U(1)_B$ was extensively studied. 
However, the color {\it variant} quantity such as diquark 
responsible for color superconductivity was not included 
in the low energy effective theory. 
Therefore, we should somehow extend their analysis 
to have diquark degrees of freedom. 
In the light of this fact, 
we propose here a SUSY model of color superconductivity 
where quarks and diquarks are dynamically generated as composites 
by a newly introduced strong gauge dynamics \cite{MT}. 

\section{Model}
Our model is based on an ${\cal N}=1$ SUSY $SO(N_c)$ gauge theory 
with $N_f(=N_c-4)$ flavors \cite{IS}. 
The flavor symmetry in the original model $SU(N_f)$ is extended 
to $SU(3)_C \times SU(N_f)_L \times SU(N_f)_R$, 
where $SU(3)_C$ is a usual color gauge group and is assumed to 
be weakly gauged compared to $SO(N_c = 6N_f + 5)$ gauge group, 
$SU(N_f)_L \times SU(N_f)_R$ is a chiral symmetry. 
Matter content of the supermultiplets is summarized below;
\begin{eqnarray}
\label{elementary}
Q &=& (\fund, \fund, \fund, {\bf 1})_{1,1,1}, \\
\bar{Q} &=& (\fund, \afund, {\bf 1}, \afund)_{-1,1,1}, \\
X &=& (\fund, {\bf 1}, {\bf 1}, {\bf 1})_{0, -6 N_f, 3 - N}
\end{eqnarray}
where the representations in the parenthesis are transformation properties 
under the group $SO(6N_f+5) \times SU(3)_C \times SU(N_f)_L \times SU(N_f)_R$. 
The numbers in the subscripts are charges for nonanomalous $U(1)$ 
global symmetries $U(1)_B \times U(1)_A \times U(1)_R$, 
which each $U(1)$ symmetry is linear combination of 
the original anomalous $U(1)$ symmetries.

$SO$ gauge theory mentioned above is known to be 
in the cofining phase \cite{IS}. 
Massless degrees of freedom describing the low energy effective theory 
below the dynamical scale $\Lambda_{SO}$ are given 
by the following gauge invariant composite superfields: 
\begin{eqnarray}
\label{comp1}
(Q^2) &=& (\afund, \asym, {\bf 1})_{2, 2, 2}, (\sym, \sym, {\bf 1})_{2,2,2}, \\
(\bar{Q}^2) &=& (\fund, {\bf 1}, \overline{\asym})_{-2, 2, 2}, 
(\overline{\sym}, {\bf 1}, \overline{\sym})_{-2,2,2}, \\
(X^2) &=& ({\bf 1}, {\bf 1}, {\bf 1})_{0, -12N_f, 6 - 2N}, \\
(QX) &=& (\fund, \fund, {\bf 1})_{1, 1 - 6N_f, 4 - N}, \\
(\bar{Q}X) &=& (\afund, {\bf 1}, \afund)_{-1, 1 - 6N_f, 4 - N}, \\
\label{comp2}
(Q\bar{Q}) &=& ({\bf 1}, \fund, \afund)_{0,2,2}, 
({\bf 8}, \fund, \afund)_{0, 2, 2}
\end{eqnarray}
where the representations in the parenthesis are 
those under the group $SU(3)_C \times SU(N_f)_L \times SU(N_f)_R$. 
The numbers in the subscripts are charges for nonanomalous $U(1)$ symmetries 
$U(1)_B \times U(1)_A \times U(1)_R$. 
Note that $Q^2(\bar{Q}^2)$ has symmetric (its conjugate) and anti-symmetric 
(its conjugate) representations 
under $SU(3)_C \times SU(N_f)_L \times SU(N_f)_R$ 
because $SO(N)$ indices are contracted 
symmetrically and the superfields are bosonic. 
One can see from the transformation properties 
that superfields with anti-symmetric representation 
in $Q^2$ and $\bar{Q}^2$ correspond to ``diquark" supermultiplet and 
$QX$ and  $\bar{Q}X$ correspond to ``quark" supermultiplet. 
Thus, quark and diquark supermultiplets are generated as composites and 
coexist in the low energy theory. 
What is more nontrivial is that these composites satisfy 
t' Hooft anomaly matching conditions, which implies that 
massless degrees of freedom are completely determined 
despite the strong coupling in the infrared below the scale $\Lambda_{SO}$.

Incorpolating the chemical potential can be regarded 
as the time component of a fictitious gauge field of 
$U(1)_B$ symmetry at zero temperature. 
For fermion field, 
it is described as 
\begin{eqnarray}
{\cal L}_{\psi} = i\bar{\psi} \partial \!\!\!/ \psi, \quad
\Delta{\cal L} = \mu \psi^\dag \psi = \mu \bar{\psi} \gamma^0 \psi 
\Rightarrow {\cal L}_{\psi} + \Delta{\cal L} 
= i\bar{\psi} (\partial \!\!\!/ -igA \!\!\!/) \psi, 
\end{eqnarray}
where $A_\nu = {\rm diag}(\mu/g,0,0,0)$, 
$g$ is the gauge coupling constant of $U(1)_B$. 
Since we are considering a SUSY theory, the corresponding argument 
for scalar fields is present, 
\begin{eqnarray}
{\cal L}_{\phi} + \Delta{\cal L} = 
(\partial^\nu -ig A^\nu)^\dag \phi^\dag (\partial_\nu - ig A_\nu) \phi. 
\end{eqnarray}
We notice that the above scalar Lagrangian with the finite chemical potential 
includes a SUSY breaking scalar mass term $+ \mu^2 \phi^\dag \phi$. 
Therefore, we impose the condition $\mu < \Lambda_{SO}$ 
since we can utilize exact results in SUSY gauge theory reliably. 
Furthermore, the SUSY breaking scalar mass squared is found to be negative, 
namely tachyon, which implies that UV theory is unstable. 
In order to avoid this situation, we have to add 
both (positive) SUSY breaking mass in the potential 
and SUSY mass in the superpotential.\footnote{The same symbols 
for the superfields and the corresponding scalar components are used.} 
\begin{eqnarray}
\Delta W_{{\rm SUSY}} &=& m_{ij} \bar{Q}_i Q_j 
\Rightarrow \Delta V_{{\rm SUSY}} = m^2 (|Q|^2 + |\bar{Q}|^2), \\
\Delta V_{{\rm SUSY~breaking}} &=& m^2_{{\rm soft}} (|Q|^2 + |\bar{Q}|^2), 
\end{eqnarray}
which results in the following UV potential 
\begin{eqnarray}
V_{{\rm UV}} &=& 
(m^2 + m^2_{{\rm soft}} - \mu^2)(|Q|^2 + |\bar{Q}|^2).
\end{eqnarray}
As will be shown later, the critical chemical potential 
where the color superconductivity happens is larger 
than the soft SUSY breaking mass. 
Thus, we need to add SUSY masses in the superpotential and 
have to impose the condition $m^2 + m^2_{{\rm soft}} - \mu^2 > 0$ 
to obtain the UV stable potential.

The low energy superpotential is generated by the gaugino condensation 
in the unbroken gauge group $SO(4) \simeq SU(2)_L \times SU(2)_R$, 
\begin{eqnarray}
W_{{\rm eff}} 
= 2(\epsilon_L + \epsilon_R)
\left( \frac{\Lambda_{SO(N)}^{6N_f+4}}{[{\rm det}(Q\bar{Q})] X} 
\right)
\end{eqnarray}
where $\epsilon_{L,R} = \pm 1$ are phase factors reflecting 
the number of SUSY vacua suggested from Witten index \cite{Witten}. 
There are two physically inequivalent branches, 
one is $W_{{\rm eff}} = 0(\epsilon_L = -\epsilon_R)$ 
and the other is $W_{{\rm eff}} \ne 0(\epsilon_L = \epsilon_R)$. 
Here, we focus on the former case throughout this article, 
which is relevant for the color supercoductivity.

In order to obtain the scalar potential, 
we need K\"ahler potential for composite fields. 
It is impossible to determine the K\"ahler potential exactly 
because of its nonholomorphicity, 
but the leading term can be fixed by using the argument of 
analytic continuation into superspace \cite{AR} in the expansion of 
the SUSY breaking scale to $\Lambda_{SO}$. 
The effective K\"ahler potential for composite fields 
is fixed by symmetries 
and the renormalization group (RG) invariance, 
\begin{eqnarray}
\label{effkahler}
K_{{\rm eff}} &=& c_{(Q^2)} \frac{{\cal Z}_{(Q^2)}}{I} (Q^2)^\dag e^{2V_B} (Q^2) 
+ c_{(\bar{Q}^2)} \frac{{\cal Z}_{(\bar{Q}^2)}}{I} 
(\bar{Q}^2)^\dag e^{-2V_B} (\bar{Q}^2) \nonumber \\
&+& c_{(X^2)} \frac{{\cal Z}_{(X^2)}}{I} (X^2)^\dag (X^2) 
+ c_{(QX)} \frac{{\cal Z}_{(QX)}}{I} (QX)^\dag e^{V_B} (QX) \nonumber \\
&+& c_{(\bar{Q}X)} \frac{{\cal Z}_{(\bar{Q}X)}}{I} 
(\bar{Q}X)^\dag e^{-V_B} (\bar{Q}X) 
+ c_{(Q\bar{Q})} \frac{{\cal Z}_{(Q\bar{Q})}}{I} (Q\bar{Q})^\dag (Q\bar{Q}), 
\end{eqnarray}
where overall coefficients $c$'s are 
of order ${\cal O}(1)$ unknown constants. 
The exponential factors for QCD are suppressed. 
$V_B$ is a background vector superfield $U(1)_B$ with a VEV 
$\langle V_B \rangle = \bar{\theta} \sigma^\mu \theta 
\langle A_\mu \rangle, \quad \langle A_\mu \rangle = (\mu/g_B,0,0,0)$. 
$g_B$ is a gauge coupling constant and 
$\mu$ is a chemical potential. 
Wave function renomalization constants $Z_i$ are promoted 
to a superfield ${\cal Z}_i$ 
\begin{equation}
{\cal Z}_i = Z_i \left[ 1 - \theta^2 \bar{\theta}^2 m_{{\rm soft}}^2 
\right]
\end{equation}
where $m_{{\rm soft}}$ is a soft SUSY breaking scalar mass in the UV 
and taken to be universal. 
The quantity $I$ is a spurious $U(1)$ symmetry and the RG invariant 
superfield, 
\begin{equation}
\label{invI}
I = \Lambda_h^\dag {\cal Z}^{2T/b_0} \Lambda_h
\end{equation}
where $T$ is the total Dynkin index of the matter fields, 
$b_0$ is the 1-loop beta function coefficient and 
$\Lambda_h = \mu_{UV} {\rm exp}[-8 \pi^2 S(\mu_{UV})/b_0], 
S(\mu_{UV}) = \frac{1}{g^2}\left( 1+ \theta^2 \frac{m_\lambda}{2}\right)
(m_\lambda: {\rm gaugino~mass})$. 
Note that a spurious $U(1)$ transformations are given by 
\begin{eqnarray}
&&Q_r \to e^A Q_r, {\cal Z}_r \to e^{A+A^\dag}{\cal Z}_r, \\
&&S(\mu_{{\rm UV}}) \to S(\mu_{{\rm UV}})-\frac{T}{4\pi^2}A
\end{eqnarray}
where $A$ is a chiral superfield.

Now, we are ready to obtain scalar masses for composites 
from the above K\"ahler potential,
\begin{eqnarray}
\label{scalar1}
\tilde{m}^2_{(Q^2)} = \tilde{m}^2_{(\bar{Q}^2)} &=& \frac{6N_f + 7}{2(3N_f + 2)} 
m_{{\rm soft}}^2 -\mu^2, \\
\label{scalar2}
\tilde{m}^2_{(X^2)} = \tilde{m}^2_{(Q\bar{Q})} &=& \frac{6N_f + 7}{2(3N_f + 2)} 
m^2_{{\rm soft}} > 0, \\
\label{scalar3}
\tilde{m}^2_{(QX)} = \tilde{m}^2_{(\bar{Q}X)} &=& \frac{6N_f + 7}{2(3N_f + 2)} 
m_{{\rm soft}}^2 - \frac{1}{4} \mu^2. 
\end{eqnarray}
One can see that $\tilde{m}^2_{(Q^2)} = \tilde{m}^2_{(\bar{Q}^2)} < 0$ 
if the chemical potential is larger than some critical value 
$\mu > \mu_* \equiv \sqrt{\frac{6N_f + 7}{2(3N_f + 2)}} m_{{\rm soft}}$. 
Thus, the diquarks condensate at the finite chemical potential, 
but the condensate is the scalar component of the diquark supermultiplet.

The above argument for the behavior of the scalar component of 
the diquark supermultiplet is valid for the vanishing superpotential case. 
For the case with nonvanishing superpotetial, on the other hand, 
it is found that the scalar potential is very complicated, 
\begin{eqnarray}
\label{potential}
V &=& \left( \frac{\partial^2 K_{{\rm eff}}}{\partial \Phi_i \partial \Phi^*_j} 
\right)^{-1} \left( \frac{\partial W_{{\rm eff}}}{\partial \Phi_i} \right)
\left( \frac{\partial W^*_{{\rm eff}}}{\partial \Phi^*_j} \right) 
\Lambda^2_{SO(N)} \nonumber \\
&+& \frac{1}{\Lambda^2_{SO(N)}} \left[
\tilde{m}^2_{(Q^2)}|(Q^2)|^2 + \tilde{m}^2_{(\bar{Q}^2)} |(\bar{Q}^2)|^2 
\right. \nonumber \\
&+& \left. \tilde{m}^2_{(X^2)} |(X^2)|^2 
+ \tilde{m}^2_{(QX)} |(QX)|^2 
+ \tilde{m}^2_{(\bar{Q}X)} |(\bar{Q}X)|^2 
+ \tilde{m}^2_{(Q\bar{Q})} |(Q\bar{Q})|^2 \right] 
\end{eqnarray}
where $\Phi_i$ denote the scalar component of composite superfields. 
$W_{{\rm eff}}$ includes SUSY mass terms. 
Therefore, we give here a qualitative discussion 
on the scalar potential behavior 
instead of an explicit minimization of the scalar potential. 
Note that F-term contributions to the scalar potential 
in the first line have a runaway behavior, 
which make the fields VEV away from the origin. 
For $\mu=0$, all SUSY breaking scalar mass squareds are positive, 
which set the fields VEV at the origin. 
Therefore all composites are expected to develop nonvanishing VEVs 
by balancing terms between the runaway potential 
and the SUSY breaking scalar mass terms. 
Even if we take into account that 
the scalar diquark mass squareds become negative for $\mu > \mu_*$, 
qualitative features of phase transition remains unchanged. 
In any case, the case of nonzero superpotential 
is irrelevant to the phase of color superconductivity of our interest. 
Even if we compare the vacuum energy in both cases, 
the case with vanishing superpotential seems to be energetically favored.

\section{Summary}
We have proposed a SUSY composite model of color superconductivity, 
which is based on a SUSY $SO(N_f+4)$ gauge theory with $N_f$ flavors. 
This model is in the confining phase in the infrared region and 
qurak and diquark supermultilets are dynamically generated 
as massless degrees of freedom by a newly introduced $SO$ 
strong gauge dynamics. 
Remarkably, massless degrees of freedom 
below the confining scale are completely determined 
since all independent composites satisfy anomaly matching conditions.

\begin{figure}
\begin{center}
\includegraphics[width=13cm,height=3cm]{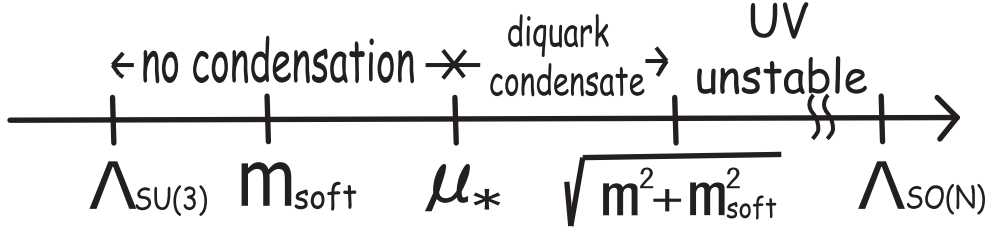}
\end{center}
\caption{\label{fig:scale} Relation among various scales are displayed. 
Horizontal axis means the energy scale. 
The possible range of the chemical potential is 
below $\sqrt{m^2 + m^2_{{\rm soft}}}$. 
There is no condensation when the chemical potential is 
between QCD scale $\Lambda_{{\rm SU(3)}}$ and 
the critical chemical potential $\mu_*$. 
The condensation of the scalar component of the diquark supermultiplet 
occurs when the chemical potential is 
between $\mu_*$ and $\sqrt{m^2 + m^2_{{\rm soft}}}$.
UV theory becomes unstable if the chemical potetntial 
is beyond the scale $\sqrt{m^2 + m^2_{{\rm soft}}}$.} 
\end{figure}

We have analytically shown that the scalar component of 
the diquark supermultiplets develop VEV at some critical chemical potential. 
The schematic picture of phase diagram is shown in Fig \ref{fig:scale}.
Although the model is not fully realistic 
in that the scalar component of diquark supermultiplet
(not diquarks themselves) condense, 
we believe that it well captures some important 
aspects of the diquark condensation behavior and helps our 
understanding for the color superconductivity in real QCD.  
If there is a certain intermediate region of the chemical potential
where quarks are deconfined but not superconducting yet,
owing to the strong quark-quark correlation, the system may be
well described by a compositon of quarks and diquarks. Then
the analysis performed in this paper will help us with
comprehending the behavior of such a system.

As future directions, 
it is interesting to extend our analysis to different flavor cases. 
In particular, Seiberg dual description \cite{Seiberg} 
might give a better understanding 
of color superconductivity. 
In order to fully understand the phase structure of QCD, 
it is necessary to take into account the finite temperature effects 
and then to study the scalar component of the diquark supermultiplet 
condensation behavior on the temperature-chemical potential plane 
as shown in Fig \ref{fig:phase}. 

\vspace*{1cm}

\begin{center}
{\bf Acknowledgments}
\end{center}
We would like to thank the organizers of the workshop for giving us
an opportunity to present this talk. 
We are supported 
by Special Postdoctoral Researchers Program at RIKEN 
(No.~A12-52040(N.M.) and No.~A12-52010(M.T.)).

\bibliographystyle{plain}

\end{document}